\documentclass[a4paper,onecolumn,aps,prl]{revtex4}
\usepackage{amsfonts}
\usepackage{latexsym}
\usepackage{euscript}
\usepackage{graphicx}
\usepackage{amsmath}
\usepackage{amsbsy}
\usepackage{amsthm}
\usepackage{bbm}
\usepackage{dsfont}
\usepackage{color}
\usepackage{hyperref}
\usepackage[symbol]{footmisc}

\newcommand{\tr}{\textnormal{Tr}}

\newcommand{\be}{\begin{equation}}
\newcommand{\ee}{\end{equation}}
\newcommand{\beq}{\begin{eqnarray}}
\newcommand{\eeq}{\end{eqnarray}}

\begin{document}

\title{Relativistic Quantum Metrology:\\Exploiting relativity to improve quantum measurement technologies}

\author{Mehdi Ahmadi$^{1}$, David Edward Bruschi$^{2}$, Carlos Sab{\'\i}n$^{1}$, Gerardo Adesso$^{1}$, and Ivette Fuentes$^{1,*}$}

\affiliation{$^{1}$School of Mathematical Sciences, University of Nottingham, University Park,
Nottingham NG7 2RD, United Kingdom\\
$^{2}$School of Electronic and Electrical Engineering, University of Leeds, Woodhouse Lane,  Leeds, LS2 9JT,  United Kingdom \\
$^*$ Corresponding author: \texttt{ivette.fuentes@nottingham.ac.uk}}

\begin{abstract}
{\bf We present a framework for relativistic quantum metrology that is useful for both Earth-based and space-based technologies. Quantum metrology has been so far successfully applied to design precision instruments such as clocks and sensors which outperform classical devices by exploiting quantum properties. There are advanced plans to implement these and other quantum technologies in space, for instance Space-QUEST and Space Optical Clock projects intend to implement quantum communications and quantum clocks at regimes where relativity starts to kick in. However, typical setups do not take into account the effects of relativity on quantum properties. To include and exploit these effects, we introduce techniques for the application of metrology to quantum field theory. Quantum field theory properly incorporates quantum theory and relativity, in particular, at regimes where space-based experiments take place. This framework allows for high precision estimation of parameters that appear in quantum field theory including proper times and accelerations.  Indeed, the techniques can be applied to develop a novel generation of  relativistic quantum technologies for gravimeters, clocks and sensors.  As an example, we present a high precision device which in principle improves the state-of-the-art in quantum accelerometers by exploiting relativistic effects.}
\end{abstract}
\maketitle

Quantum technologies are widely expected to bring about many key technological advances this century. Experiments in quantum communication are rapidly progressing from table-top to space-based setups. For instance, in 2012 a teleportation protocol was successfully performed across a distance of 143km by the group led by A. Zeilinger~\cite{zeilingerteleport}. Partly motivated by this success, major space agencies, e.g., in Europe and Canada, have invested resources for the implementation of space-based quantum technologies~\cite{spaceexp1,spaceexp2,spaceexp3}. There are advanced plans to use satellites to distribute entanglement for quantum cryptography and teleportation (e.g., the Space-QUEST project~\cite{UrsinetalSpaceQuest2008}) and to install quantum clocks in space (e.g., the Space Optical Clock project~\cite{SchillerEtalSpaceOpticalClock2012}). However, at these scales relativistic effects become observable. General relativity provides an effective description of the Universe at large length scales\textemdash  observable effects can thus be expected at the regimes where satellites operate. For instance, the Global Positioning System (GPS), a system of satellites used for time dissemination and navigation, requires relativistic corrections to determine time and positions accurately~\cite{AshbyGPS2003}. Cutting-edge experiments are reaching relativistic regimes, yet the effects of gravity and motion on quantum technologies are largely unknown.
\begin{figure}[th]
\includegraphics[width=0.9\linewidth]{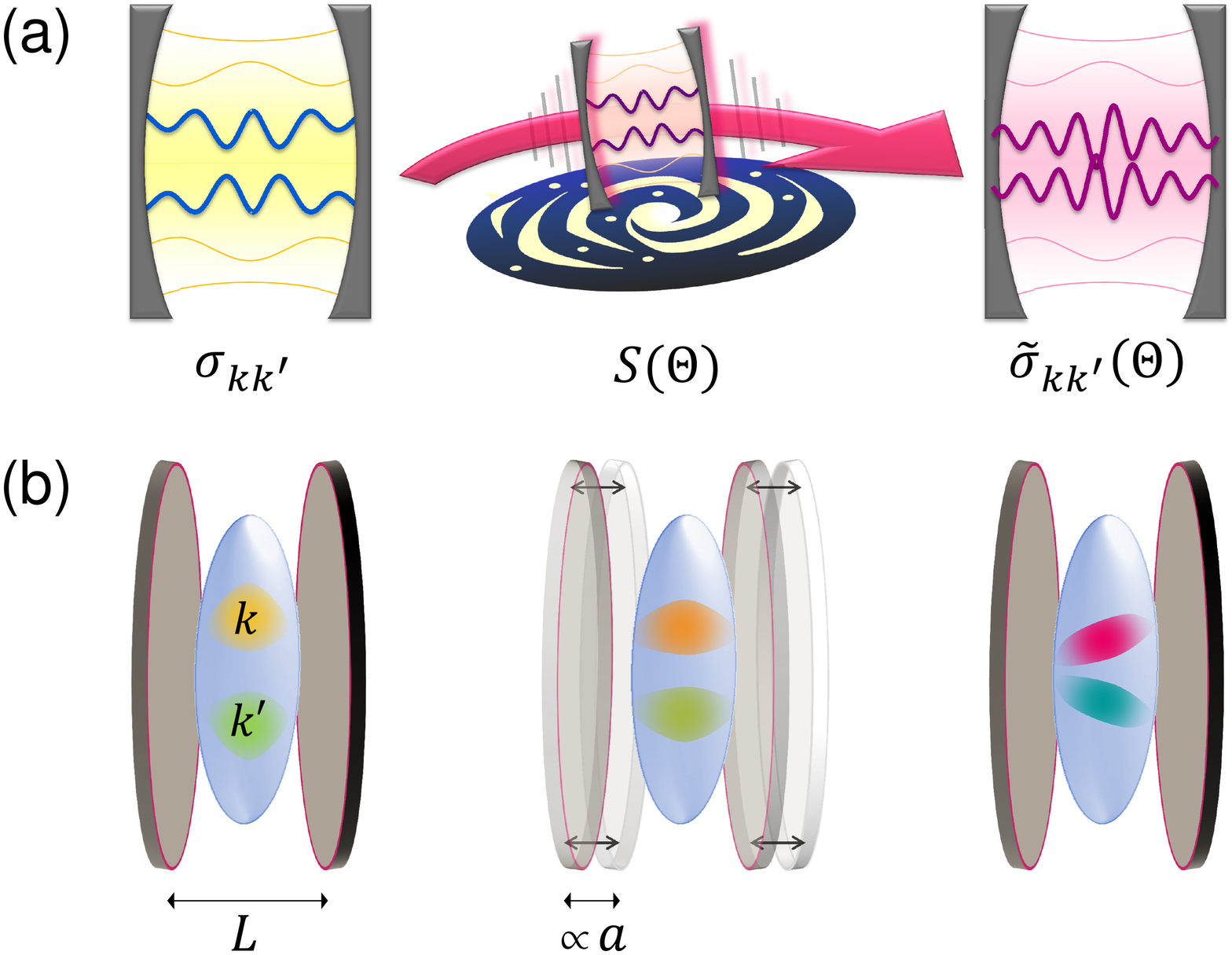}
\caption{(a) General cavity framework: the initial state of two modes of a quantum field inside a cavity, represented initially by the covariance matrix $\sigma_{kk^{\prime}}$ undergoes a relativistic transformation which depends on some parameter~$\Theta$. The transformed covariance matrix~$\tilde{\sigma}_{kk^{\prime}}$ depends on the parameter~$\Theta$, which can be estimated using quantum metrology tools. (b) Example: measurement of the acceleration in a BEC setup.} \label{fig:sketch}
\end{figure}

The inability to unify quantum theory and general relativity remains one of the biggest unsolved problems in physics today. Understanding general relativity at small length scales where quantum effects become relevant is a highly non-trivial endeavour that suffers from a scarcity of experimental guidance. Therefore, an alternative approach is to study quantum effects at large scales where experiments promise to be within reach in the near future~\cite{rideout, ursin}. However, in order to exploit quantum properties to measure position and time at scales where relativity becomes relevant, it is necessary to work within quantum field theory  which provides a description of quantum fields in curved space-time~\cite{BirrellDavies:QFbook}. It is a semiclassical description in the sense that matter and radiation are quantized but the space-time is classical. As a first step in this direction, it was shown that quantum metrology techniques can be applied to measure the Unruh effect at accelerations that are within experimental reach~\cite{aspachs,HoslerKok2013,Hosler:PhDThesis2013} and to estimate the curvature of space-time~\cite{downes}. It was also shown that entanglement can be used to determine space-time parameters such as the expansion rate of the universe~\cite{fredericivy}. An appropriate quantum field theory approach that includes the effects of quantum theory and relativity will enable the development of a new generation of quantum technologies for space. Indeed, previous work on relativistic quantum information has already addressed fundamental questions about entanglement in quantum field theory where results show that gravity, motion and space-time dynamics can create and degrade entanglement~\cite{ivyalsingreview}. Recent work~\cite{teleportationico} shows that this implies that relativistic motion produces observable effects on quantum communication. These preliminary results are of great importance for the space-based quantum experiments that will test quantum theory at large scales. 

In this paper we develop a new framework for relativistic quantum metrology by incorporating and exploiting relativistic effects in quantum parameter estimation. The framework provides the necessary methodology that will enable the design and production of new measurement instruments. As an example we present an accelerometer (see Fig.~\ref{fig:sketch}) with a precision that, in principle, improves the state-of-the-art in cutting-edge experiments to detect gravity anomalies in outer space~\cite{sagas,compact}. Our device is based on the fact that acceleration produces observable relativistic effects on Bose-Einstein Condensates (BEC)~\cite{salelites}. The  motion-induced transformation of the state of the relativistic phononic field on the BEC allows an extremely accurate estimation of the magnitude of the acceleration. We use quantum metrology tools, in particular, quantum Fisher information \cite{paris} to compute optimal bounds to the error of the estimation procedure.

\section{Methods}
\subsection{Quantum field theory and the covariance matrix formalism}

We are interested in applying metrology techniques to determine with high precision parameters that appear in quantum field theory, for instance accelerations, gravitational field strengths, and proper times. In order to do so, we begin by reviewing basic concepts from quantum field theory and the covariance matrix formalism. We consider a real, scalar quantum field that obeys the Klein-Gordon equation in curved spacetime.  It is convenient to expand the field in a discrete set of modes $\{\phi_{n}|n=1,2,3,...\}$ \cite{footnote} so that we can represent the field and its transformations in the covariance matrix formalism which is commonly employed in continuous variable quantum information and quantum metrology. We refer the readers to~\cite{ourreview,nicoivy} for further details. The functions  $\phi_{n}$ are solutions to the field equation and form a complete set of orthonormal modes with respect to a chosen inner product~\cite{BirrellDavies:QFbook} denoted by $(\,.\,,\,.\,)$. The creation and annihilation operators associated to the field modes satisfy the commutation relations $[a_{m},a_{n}]=[a^{\dag}_{m},a^{\dag}_{n}]=0$ and $[a_{m},a^{\dag}_{n}]=\delta_{mn}$. The vacuum state is defined as the state that is annihilated by the operators $a_{n}$ for all~$n$, i.e., $a_{n} |0\rangle=0$.   A coordinate transformation between different observers, for example, between inertial and accelerated observers~\cite{alphacentauri}, results in a Bogoliubov transformation between $\phi_{n}$ and mode solutions in the new coordinate system denoted by  $\tilde{\phi}_{n}$. The operators associated with $\tilde{\phi}_{n}$ are $\tilde{a}_{n}$ .

The most general linear transformation between the field operators $a_{m}$ and $\tilde{a}_{n}$ is given by,
\be\label{Bogoliubov transformation}
\tilde{a}_{m}=\sum_{n} \bigl(\alpha^{*}_{mn}a_{n}+\beta^{*}_{mn}a^{\dag}_{n}\bigr)\,,
\ee
where $\alpha_{mn}=(\tilde{\phi}_{m},\phi_{n})$ and $\beta_{mn}=-(\tilde{\phi}_{n},\phi^{*}_{m})$ are the Bogoliubov coefficients. The transformed vacuum $|\tilde{0}\rangle$ is annihilated by the new annihilation operators $\tilde{a}_{n}$ for all~$n$. Note that $|\tilde{0}\rangle$ is annihilated by the initial field operators $a_n$ only if all coefficients $\beta_{mn}$ are zero. Indeed particle production manifests when $\beta_{mn}\neq0$. This occurs, for instance, in the Unruh effect where the inertial vacuum state is seen as a thermal state by uniformly accelerated observers~\cite{BirrellDavies:QFbook}. Another example of interest is that of a cavity in non-inertial motion. The vacuum state of an inertial cavity becomes populated by particles after the cavity undergoes non-uniformly accelerated motion~\cite{casimirwilson}.

Let us now consider the covariance matrix formalism, which has been very useful to investigate entanglement in quantum field theory~\cite{givyericsson,nicoivy,gcqg}. In this phase space formalism, for Gaussian states of a bosonic field, all the relevant information about the state is encoded in the first and second moments of the field. In particular, the second moments are described by the covariance matrix $\sigma_{ij}=\langle X_{i} X_{j}+X_{j}X_{i}\rangle-2\langle X_{i}\rangle\langle X_{j}\rangle$, where $\langle\,.\,\rangle$ denotes the expectation value and the quadrature operators $X_{i}$ are the generalized position and momentum operators of the field modes. In this paper we follow the conventions used in~\cite{ourreview,nicoivy}, i.e., the operators for the $n$-th mode are given by $X_{2n-1}=\frac{1}{\sqrt{2}}(a_{n}+a^{\dag}_{n})$ and $X_{2n}=\frac{-i}{\sqrt{2}}(a_{n}-a^{\dag}_{n})$. The covariance matrix formalism enables elegant and simplified calculations and has been proven useful to define and analyze computable measures of bipartite and multipartite entanglement for Gaussian states~\cite{ourreview,gcqg}.

Every unitary transformation in Hilbert space that is generated by a quadratic Hamiltonian can be represented as a symplectic matrix $S$ in phase space. These transformations form the real symplectic group $Sp(2n,\mathds{R})
$, the group of real $(2n\times2n)$ matrices that leave the symplectic form $\Omega$ invariant, i.e., $S\Omega S^{T}=\Omega$, where $\Omega=\bigoplus_{i=1}^{n}\Omega_i$ and $\Omega_i=\left(
            \begin{array}{cc}
              0 & 1 \\
              -1 & 0 \\
            \end{array}
          \right)
$\,. The time evolution of the field, as well as the Bogoliubov transformations, can be encoded in this structure. The symplectic matrix corresponding to the Bogoliubov transformation in Eq.~(\ref{Bogoliubov transformation}) can be written in terms of the Bogoliubov coefficients as
\be\label{Bogosymplectic}
S=\left(
  \begin{array}{cccc}
    \mathcal{M}_{11} & \mathcal{M}_{12} & \mathcal{M}_{13} & \cdots \\
    \mathcal{M}_{21} & \mathcal{M}_{22} & \mathcal{M}_{23} & \cdots \\
    \mathcal{M}_{31} & \mathcal{M}_{32} & \mathcal{M}_{33} & \cdots \\
    \vdots & \vdots & \vdots & \ddots
  \end{array}
\right)\,,
\ee
where the $\mathcal{M}_{mn}$ are the $2\times2$ matrices
\be\label{Mmatrices}
\mathcal{M}_{mn}=\left(
                   \begin{array}{cc}
                     \mathrm{Re}(\alpha_{mn}-\beta_{mn}) & \mathrm{Im}(\alpha_{mn}+\beta_{mn}) \\
                     -\mathrm{Im}(\alpha_{mn}-\beta_{mn}) & \mathrm{Re}(\alpha_{mn}+\beta_{mn})
                   \end{array}
                 \right)\,.
\ee
Here $\mathrm{Re}$ and $\mathrm{Im}$ denote the real and imaginary parts, respectively. The covariance matrix after a Bogoliubov transformation is given by $\tilde{\sigma}=S\sigma S^{T}$. Let us suppose that we are only interested in the state of two modes~$k$ and~$k^{\prime}$ after the transformation. A great advantage of the covariance matrix formalism is that the trace operation over a mode is implemented simply by deleting the row and column corresponding to that mode. Consider that the initial state of the quantum field is a general Gaussian state for modes~$k$ and~$k^{\prime}$, and that all other modes are in their vacuum state.  The covariance matrix for modes $k$ and~$k^{\prime}$ is given by
\be\label{initialcm}
\sigma_{kk^{\prime}}=\left(
               \begin{array}{cc}
                 \psi_k & \phi_{kk^{\prime}} \\
                 \phi^{T}_{kk^{\prime}} & \psi_k^{\prime}
               \end{array}
             \right)\,,
\ee
where $\psi_{k}$, $\psi_{k^{\prime}}$ and $\phi_{kk^{\prime}}$ are $2\times2$ matrices. Then the transformed covariance matrix is given by
\be\label{transformedCM}
 \tilde{\sigma}_{kk^{\prime}}=\left(
                        \begin{array}{cc}
                          C_{kk} & C_{kk^{\prime}} \\
                          C_{k^{\prime}k} & C_{k^{\prime}k^{\prime}} \\
                        \end{array}
                      \right)\,,
\ee
where
\begin{equation}
 C_{ij} =\,\mathcal{M}_{ki}^T \psi _k \mathcal{M}_{kj}\,
    +\,\mathcal{M}_{k^{\prime}i}^{T}\phi^{T}_{kk^{\prime}}\mathcal{M}_{kj}\,
    +\,\mathcal{M}_{ki}^{T}\phi ^{T}_{kk^{\prime}}\mathcal{M}_{k^{\prime}j} +\,\mathcal{M}_{k^{\prime}i}^{T}\psi_{k^{\prime}}\mathcal{M}_{k^{\prime}j}\, +\, \sum _{ n\neq i,j}\mathcal{M}_{ni}^{T} \mathcal{M}_{nj}\,.
 \label{Cij}
\end{equation}

\subsection{Metrology techniques}

Having written the state of the field and its transformations in the covariance matrix formalism we are now ready to apply metrology techniques \cite{paris} that have been developed for continuous variable systems. In this section we briefly review some basic tools \cite{advances,pinel,qmqft}. The aim in quantum metrology is to provide a strategy to determine the value of a parameter $\Theta$ with high precision when the parameter is not an observable of the system. Temperature, time, acceleration, and coupling strengths are good examples. A strategy corresponds to finding optimal initial states and measurements on the final states.
In order to estimate the parameter with high precision it is necessary to distinguish two states $\rho_{\Theta}$ and $\rho_{\Theta+d\Theta}$ that differ by an infinitesimal change $d\Theta$ of the parameter $\Theta$. The operational measure that quantifies the distinguishability of these two states is the Fisher information \cite{paris}. Let us suppose that an experimenter performs $N$ independent measurements to obtain an unbiased estimator $\hat{\Theta}$ for the parameter $\Theta$. The Fisher Information $F(\Theta)$ gives a lower bound to the  mean-square error via the classical Cram\'er-Rao inequality~\cite{Cramer:Methods1946}, i.e., $\langle (\Delta \hat{\Theta})^{2}\rangle\geq\frac{1}{NF(\Theta)}$, where $F(\Theta)=\int\!d\lambda~p(\lambda|\Theta) (d\,\ln [p(\lambda|\Theta)]/d\lambda)^{2}$ and $p(\lambda|\Theta)$ is the likelihood function with respect to a chosen positive operator-valued measurement (POVM)~$\{\hat{O}_{\lambda}\}$ with $\sum_{\lambda}\hat{O}_{\lambda}=\mathds{1}$\,.
Optimizing over all the possible quantum measurements provides an even stronger lower bound~\cite{BraunsteinCaves1994}, i.e.,
\be\label{Cramer-Rao}
N\langle (\Delta \hat{\Theta})^{2}\rangle\geq\frac{1}{F(\Theta)}\geq \frac{1}{H(\Theta)},
\ee
where $H(\Theta)$ is the quantum Fisher information (QFI). This quantity is obtained by determining the eigenstates of the symmetric logarithmic derivative $\Lambda_{\rho_\Theta}$ defined by  $2\frac{d\rho_{\Theta}}{d\Theta}=\Lambda_{\rho_{\Theta}} \rho_{\Theta}+\rho_{\Theta} \Lambda_{\rho_{\Theta}}\,$. Alternatively, the QFI can be related to the Uhlmann fidelity~$\mathcal{F}$ of the two states $\rho_{\Theta}$ and $\rho_{\Theta+d\Theta}$ through
 \begin{equation}
 H(\Theta)=\frac{8\left[1-\sqrt{\mathcal{F}(\rho_{\Theta},\rho_{\Theta+d\Theta})}\right]}{d\Theta^{2}}, \label{quantumfishinfo}
\end{equation}
where $\mathcal{F}(\rho_1,\rho_2)=(\tr\sqrt{\sqrt{\rho_1}\rho_2\sqrt{\rho_1}})^{2}\,$. The optimal POVMs for which the quantum Cram\'er-Rao bound~(\ref{Cramer-Rao}) becomes asymptotically tight can be computed from $\Lambda_{\rho_\Theta}$ \cite{monras}. Unfortunately, these optimal measurements are usually not easily implementable in the laboratory. Nevertheless, in typical problems involving optimal implementations one can devise suboptimal strategies involving feasible measurements such as homodyne or heterodyne detection, see, e.g.,~\cite{advances}. Here we are interested in assessing metrology strategies based on the quantum Cram\'er-Rao bound and thus on the QFI, since our aim is to investigate how well one can in principle determine a parameter that appears in quantum field theory. As an example we will consider acceleration and we will show how our technique can be applied to develop a quantum accelerometer that exploits relativistic effects. By doing so, it is in principle possible to improve the state-of-the-art in accelerometers. Before presenting our example we return to our general discussion.\\

We consider a bosonic quantum field which undergoes a $\Theta$-dependent Bogoliubov transformation, where $\Theta$ is the parameter we want to estimate. For example, the transformation could be the expansion of the universe and the parameter in this case is the expansion rate. We assume that the initial state of the field is given by Eq.~(\ref{initialcm}). To estimate $\Theta$ we must calculate the fidelity $\mathcal{F}(\tilde\sigma_{kk^{\prime}}[\Theta],\tilde\sigma_{kk^{\prime}}[\Theta+d\Theta])$, where  the transformed state $\tilde\sigma_{kk^{\prime}}[\Theta]$ is given by Eq.~(\ref{transformedCM}). If $\tilde\sigma_{kk^{\prime}}$ is a two-mode Gaussian state with zero initial first moments, the fidelity is given by
\cite{MarianMarian}
\begin{eqnarray}
    \mathcal{F}(\tilde\sigma_{kk^{\prime}}[\Theta],\tilde\sigma_{kk^{\prime}}[\Theta+d\Theta])\,&=\, \label{MarianMariantwomodeGaussianFidelity}
    \displaystyle\frac{1}{\Pi[\Theta, \Theta+d\Theta]+\sqrt{\Pi[\Theta, \Theta+d\Theta]^{2}-\Delta[\Theta, \Theta+d\Theta]}}\,,
\end{eqnarray}
where $\Pi[\Theta, \Theta+d\Theta]=\sqrt{\Gamma[\Theta, \Theta+d\Theta]}+\sqrt{\Lambda[\Theta]\Lambda[\Theta+d\Theta]}\,$, and
\label{matrices}
\begin{eqnarray}
\Gamma[\Theta, \Theta+d\Theta] &=& \frac{1}{16}\det(\Omega\,\tilde{\sigma}_{kk^{\prime}}[\Theta]\,\Omega \,\tilde{\sigma}_{kk^{\prime}}[\Theta+d\Theta]\,-\,\mathds{1})\,,\\
\Lambda[\Theta] &=& \frac{1}{4} \det(\tilde\sigma_{kk^{\prime}}[\Theta]\,+\,i\,\Omega)\,,\\
\Delta [\Theta, \Theta+d\Theta] &=& \frac{1}{16}\det(\tilde\sigma_{kk^{\prime}}[\Theta]\,+\,\tilde\sigma_{kk^{\prime}}[\Theta+d\Theta])\,,
\end{eqnarray}
where $\mathds{1}$ is the identity matrix. Note that we follow the conventions used in~\cite{ourreview,nicoivy} for the normalization of the covariance matrix, which differ from other conventions~\cite{MarianMarian}. 

We now present an application of the techniques to measure accelerations.

\section{Results: Measuring acceleration}

We consider a bosonic quantum field in flat  spacetime confined in a cavity undergoing non-uniform motion.  The motion-induced transformation of the quantum field depends on the magnitude of the acceleration. We will show that, using the techniques explained above, this fact can be used as the working principle of a highly accurate and sensitive accelerometer.

Our aim is to determine the precision with which the acceleration can be estimated from measurements on the field modes. We present our analysis in $(1+1)$-dimensional spacetime with metric signature $(-+)$. Additional transverse dimensions can be included via their contribution to the effective field mass as discussed in~\cite{FriisLeeLouko2013,alphacentauri}. We consider Dirichlet boundary conditions at the cavity walls.  The details of the chosen boundary condition slightly modify the quantitative features of the model~\cite{FriisLeeLouko2013}, but not qualitatively. Physical implementations of this setup can be realised by optical cavities~\cite{alphar}, superconducting circuits~\cite{teleportationico}, or Bose-Einstein condensates~\cite{salelites}.

The cavity is considered to be initially at rest and the state of the field given by Eq.~(\ref{initialcm}). After a general trajectory the field modes undergo a Bogoliubov transformation. In order to treat the problem analytically we assume that the Bogoliubov coefficients which relate the initial and final state of the field have a series expansion in terms of a dimensionless parameter~$h$, such that
\begin{eqnarray} \label{Maclaurin expansion}
\alpha_{mn} &=&\,    \alpha^{\raisebox{0.5pt}{\tiny{$\,(0)$}}}_{mn}\,+\,
                    \alpha^{\raisebox{0.5pt}{\tiny{$\,(1)$}}}_{mn}\,h\,+\,
                    \alpha^{\raisebox{0.5pt}{\tiny{$\,(2)$}}}_{mn}\,h^{2}\,+\,O(h^{3})\,,
                    \label{Maclaurin expansion alphas}\\
\beta_{mn}  &=&\,    \beta^{\raisebox{0.5pt}{\tiny{$\,(1)$}}}_{mn}\,h\,+\,
                    \beta^{\raisebox{0.5pt}{\tiny{$\,(2)$}}}_{mn}\,h^{2}\,+\,O(h^{3})\,,
                    \label{Maclaurin expansion betas}
\end{eqnarray}
where
\begin{equation}
h=\frac{a\,L}{c_s^{2}}=\frac{a\,L\,n^2}{c^2}.\label{eq:h}
\end{equation}
Here $a$ is the proper acceleration of the cavity at its center,~$L$ is the length of the cavity in its instantaneous
rest frame, $c_s$ is the propagation speed of the excitations of the quantum field inside the cavity and we introduce $n=c/c_s$, where $c$ is the speed of light in vacuum. In general the motion of the cavity can be an arbitrary combination of segments of uniform acceleration and inertial motion.
For example, we can consider a finite period of uniform acceleration, a repetition of identical trajectory segments, or even sinusoidal oscillation with a fixed amplitude~\cite{alphar}.

 The coefficients $\alpha^{\raisebox{0.5pt}{\tiny{$\,(0)$}}}_{mn}=G_{m}\delta_{mn}$, where $|G_{m}|=1$, are the phases accumulated during both uniform acceleration and inertial segments. Furthermore, the first order coefficients are zero on the diagonal, i.e., $\alpha^{\raisebox{0.5pt}{\tiny{$\,(1)$}}}_{nn}=\beta^{\raisebox{0.5pt}{\tiny{$\,(1)$}}}_{nn}=0$. In this paper we suppose that the two modes~$k$ and~$k^{\prime}$ have opposite parity, $(k-k^{\prime})$ is odd. This extra assumption causes the coefficients $\alpha_{kk^{\prime}}$ and $\beta_{kk^{\prime}}$ to contain only odd powers of~$h$ in their series expansions.

Let us consider a particular initial state of the modes~$k$ and~$k^{\prime}$. We assume the state to be two single-mode squeezed states in a product form. The $2\times2$ matrices in~(\ref{initialcm}) are then $\psi_{k}=\left(\begin{array}{cc} e^{2r_k} & 0 \\ 0 & e^{-2r_k} \end{array}\right)\,$,
$\psi_{k'}=\left(\begin{array}{cc} e^{2r_{k^{\prime}}} & 0 \\ 0 & e^{-2r_{k^{\prime}}} \end{array}\right)$ and $\phi_{kk^{\prime}}=\left(\begin{array}{cc} 0 & 0 \\ 0 & 0 \end{array}\right)$.

Currently, we are also investigating other initial states such as entangled Gaussian and Fock states. In this paper we restrict our analysis to the state mentioned above which already produces positive results. The transformed covariance matrix~$\tilde{\sigma}_{kk^{\prime}}(h)$ is obtained using Eqs.~(\ref{Mmatrices}),~(\ref{transformedCM}), ~(\ref{Cij}) and ~(\ref{Maclaurin expansion}), with our particular initial state. The QFI is given by
\begin{equation}
    H(h)=\frac{8\left[1-\sqrt{\mathcal{F}(\tilde{\sigma}_{kk^{\prime}}(h),\tilde{\sigma}_{kk^{\prime}}(h+dh)}\right]}{dh^{2}}\,,
    \label{quantumfishinfoh}
\end{equation}
where $\mathcal{F}(\tilde{\sigma}_{kk^{\prime}}(h),\tilde{\sigma}_{kk^{\prime}}(h+dh))$ can be computed using Eq.~(\ref{MarianMariantwomodeGaussianFidelity}). Using the Bloch-Messiah reduction~\cite[p.~9]{reviewGaussian} it is possible to show that $\Delta=\Gamma + \mathcal{O} (h^3, dh^3)$ and $\Lambda= \mathcal{O}(h^4\,(h+dh)^4)$. Therefore, the fidelity is given by $\mathcal{F}(\tilde{\sigma}_{kk^{\prime}}(h),\tilde{\sigma}_{kk^{\prime}}(h+dh))=1-\frac{1}{2}\,\tilde{\Gamma}\,dh^{2}$, where~$\tilde{\Gamma}$ is the term proportional to $dh^{2}$ in the Taylor expansion of~$\Gamma$. Assuming for the sake of simplicity that both modes have the same squeezing parameter $r_{k}=r_{k^{\prime}}=r$, we obtain the QFI $H(h)=H^{(0)}+ H^{(2)}\,h^2$, where
\begin{eqnarray}
H^{\raisebox{0.5pt}{\tiny{$\,(0)$}}}  &=&\Re\bigg[4 \cosh r(f^{k}_{\alpha}+f^{k}_{\beta}+f^{k'}_{\alpha}+f^{k'}_{\beta})+ 4\cosh^2r(|\alpha_{kk'}^{\raisebox{0.5pt}{\tiny{$\,(1)$}}}|^2+|\beta_{kk'}^{\raisebox{0.5pt}{\tiny{$\,(1)$}}}|^2)
-4\cosh^4r|\beta_{kk'}^{\raisebox{0.5pt}{\tiny{$\,(1)$}}}|^2\nonumber\\
&-&4\sinh^2r(G_{k'}^{*2}{\alpha_{kk'}^{\raisebox{0.5pt}{\tiny{$\,(1)$}}}}^{2}
+G_{k'}^{2}{\beta_{kk'}^{\raisebox{0.5pt}{\tiny{$\,(1)$}}}}^{2}-f^{k}_{\alpha}+f^{k}_{\beta}-f^{k'}_{\alpha}+f^{k'}_{\beta}-|\alpha_{kk'}^{\raisebox{0.5pt}{\tiny{$\,(1)$}}}|^2+|\beta_{kk'}^{\raisebox{0.5pt}{\tiny{$\,(1)$}}}|^2)\nonumber\\
&-&2\sinh2r(2\alpha_{kk'}^{\raisebox{0.5pt}{\tiny{$\,(1)$}}}\beta_{kk'}^{\raisebox{0.5pt}{\tiny{$\,(1)$}}}+2\alpha_{k'k}^{\raisebox{0.5pt}{\tiny{$\,(1)$}}}\beta_{k'k}^{\raisebox{0.5pt}{\tiny{$\,(1)$}}}-\cos(\phi_{k})\left(-f^{k}_{\alpha}+f^{k}_{\beta}-\frac{|\alpha_{kk'}^{\raisebox{0.5pt}{\tiny{$\,(1)$}}}|^2}{2}+\frac{|\beta_{kk'}^{\raisebox{0.5pt}{\tiny{$\,(1)$}}}|^2}{2}\right)
\nonumber\\&-&\cos(\phi_{k'})\left(-f^{k'}_{\alpha}+f^{k'}_{\beta}-\frac{|\alpha_{kk'}^{\raisebox{0.5pt}{\tiny{$\,(1)$}}}|^2}{2}+\frac{|\beta_{kk'}^{\raisebox{0.5pt}{\tiny{$\,(1)$}}}|^2}{2}\right)
+ 4\sinh r(G_{k}^{*2}\mathcal{G}^{\alpha \beta}_{kk}+G_{k'}^{*2}\mathcal{G}^{\alpha \beta}_{k'k'})\nonumber\\
&+&4\sinh2r\cosh^2r(\alpha_{kk'}^{\raisebox{0.5pt}{\tiny{$\,(1)$}}}\beta_{kk'}^{\raisebox{0.5pt}{\tiny{$\,(1)$}}}+\alpha_{k'k}^{\raisebox{0.5pt}{\tiny{$\,(1)$}}}\beta_{k'k}^{\raisebox{0.5pt}{\tiny{$\,(1)$}}})+2\sinh^4r(|\alpha_{kk'}^{\raisebox{0.5pt}{\tiny{$\,(1)$}}}|^2-|\beta_{kk'}^{\raisebox{0.5pt}{\tiny{$\,(1)$}}}|^2-{G_{k'}^{*}}^{2}{\alpha_{kk'}^{\raisebox{0.5pt}{\tiny{$\,(1)$}}}}^{2}-\nonumber\\
&&-G_{k'}^{2}{\beta_{kk'}^{\raisebox{0.5pt}{\tiny{$\,(1)$}}}}^{2})-\frac{1}{2}\sinh^22r(|\alpha_{kk'}^{\raisebox{0.5pt}{\tiny{$\,(1)$}}}|^2-3|\beta_{kk'}^{\raisebox{0.5pt}{\tiny{$\,(1)$}}}|^2
-{G_{k'}^{*}}^{2}{\alpha_{kk'}^{\raisebox{0.5pt}{\tiny{$\,(1)$}}}}^{2}-G_{k'}^{2}{\beta_{kk'}^{\raisebox{0.5pt}{\tiny{$\,(1)$}}}}^{2})\bigg],
        \label{fishresult}
   \end{eqnarray}
where $f_{\alpha}^{i}=\,\sum_{n\neq k,k'}|\alpha_{ni}^{\raisebox{0.5pt}{\tiny{$\,(1)$}}}|^2,
f_{\beta}^{i}=\,\sum_{n\neq k,k'}|\beta_{ni}^{\raisebox{0.5pt}{\tiny{$\,(1)$}}}|^2$ and
$\mathcal{G}^{\alpha\beta}_{ij}=\,\sum_{n\neq k,k'} \alpha_{ni}^{\raisebox{0.5pt}{\tiny{$\,(1)$}}}{\beta_{nj}^{\raisebox{0.5pt}{\tiny{$\,(1)$}}}}^{*}$. The particular form of the Bogoliubov coefficients depends on the trajectory followed by the cavity. Arbitrary trajectories composed of discrete intervals of accelerated and inertial motion were considered, for instance in~\cite{alphacentauri}, while continuous motion was addressed in~\cite{alphar}. The latter case \textemdash\ continuous, sinusoidal motion with small amplitude \textemdash\ features two different kinds of resonances. If the frequency of the oscillation matches the sum of the frequencies of the two oddly separated modes the corresponding $|\beta^{\raisebox{0.5pt}{\tiny{$\,(1)$}}}_{kk^{\prime}}|$ grows linearly with the duration of the oscillation. This gives rise to a resonant particle creation phenomenon known as the dynamical Casimir effect~\cite{casimirwilson}. If, on the other hand, the oscillation frequency is equal to the difference between the frequencies of two oddly separated modes the resonance is associated to the coefficient $|\alpha^{\raisebox{0.5pt}{\tiny{$\,(1)$}}}_{kk^{\prime}}|$ (see~\cite{alphar}). In this paper, we take advantage of these resonances to increase the QFI in Eq.~(\ref{fishresult}). Notice that our perturbative approach is restricted by the condition $H^{\raisebox{0.5pt}{\tiny{$\,(0)$}}}\,h^{2}\ll1$.

Finally, using  Eq.~(\ref{Cramer-Rao}) the optimal bound to the error in the estimation of the parameter~$h$ after $N$ measurements is obtained as $\langle\Delta\,h\rangle\geq\frac{1}{\sqrt{N\,H^{\raisebox{0.5pt}{\tiny{$\,(0)$}}}}}$. Assuming good control over the parameters~$L$ and~$c_s$, the final error in the estimation of the acceleration is just re-scaled by a factor of $c_s^{2}/L=c^2/(n^2\,L)$, that is:
\begin{equation}
\langle\Delta a\rangle\geq\frac{c^2}{n^2\,L\,\sqrt{N\,H^{\raisebox{0.5pt}{\tiny{$\,(0)$}}}}}
\end{equation}

Now we will consider a specific experimental implementation using a quasi one-dimensional BEC \cite{1dbec} on a flat spacetime metric \cite{matt,liberati} with hard-wall boundary conditions \cite{salelites,condensatebox1, condensatebox2,condensatebox3}. In the dilute regime, the BEC can be described by a mean field density plus phase fluctuations $\hat{\phi}$, which we expand in terms of the so-called Bogoliubov modes.  The modes with frequencies well below the frequencies associated to the healing length of the condensate obey a massless Klein-Gordon equation in an effective curved spacetime metric. The effective metric depends on the real spacetime metric, the background pressure $p$, energy density $\rho$, number density $n$ and flow velocity $v$ \cite{matt,liberati}. In the absence of background flows $v=0$ and for constant density (i.e. the homogenous case), the effective metric is also flat \cite{analoguegravity}.

Let us now describe the BEC after it undergoes acceleration, a situation that has been previously considered \cite{acceleratedbec}. In the comoving frame, the effective metric remains flat as long as the accelerations are small enough. In this case, we can ensure that the classical background is not excited and that  the condensate remains approximately homogenous. In \cite{acceleratedbec} it has been shown that squeezing of the Bogoliubov modes occurs when a BEC with hard-wall boundary conditions undergoes small accelerations. The density of the BEC can in principle become inhomogeneous, however, these effects are negligible in the regimes considered in our discussion. Under the circumstances mentioned above, we obtain a Klein-Gordon equation for the Bogoliubov modes of the BEC with an effective flat metric and propagation speed $c^2_s=c^2\partial p/\partial\rho$.  The hard-wall boundary conditions give rise to a mode spectrum given by $\omega_n=2\pi\times \frac{n\,c_s}{L}$,  where $L$ is the length of the condensate. Therefore, the techniques presented in the previous sections are directly applicable.  A more detailed presentation of our BEC setup can be found in \cite{salelites}.

We have chosen to use a BEC to take advantage of amplification effects due to small propagation speeds $c_s$ among other convenient experimental features that we will discuss below.  At this point, a comment on the relativistic nature of phonon production on a BEC is in order.
In the Newtonian limit where $v,c_s \ll c$ one can assume that $c_s$ is independent of $c$ as explained in more detail in reference \cite{matt}. In this limit an accelerated observer and an inertial observer are no longer related through Rindler transformations (i.e. Lorentz transformations in the uniformly accelerated case) but instead by Galilean transformations. In this case the time coordinate in the accelerated frame coincides with the time coordinate in the inertial frame since time dilation is negligible. As a consequence the cavity length remains constant in all frames and the vacua of the cavity at rest and the cavity undergoing uniform acceleration coincide. There is no particle creation, confirming that the effect we discuss is purely relativistic. In the case that $v/c$ and $c_s/c$  are not negligible, it is also possible to determine whether an effect is relativistic or not.
In this case an effect is considered to be relativistic if it disappears in the limit $c^2\rightarrow\infty$. By this criterion, photon production due to the motion of a boundary is a relativistic effect. In the BEC the excitations propagate much slower. However, the speed of propagation is directly proportional to the speed of light through $c^2_s=c^2\partial p/\partial\rho$ \cite{matt,liberati}. Therefore, the effect also disappears in the limit $c^2\rightarrow\infty$ and we therefore consider it to be relativistic.

\begin{figure}[th]
\includegraphics[width=\linewidth]{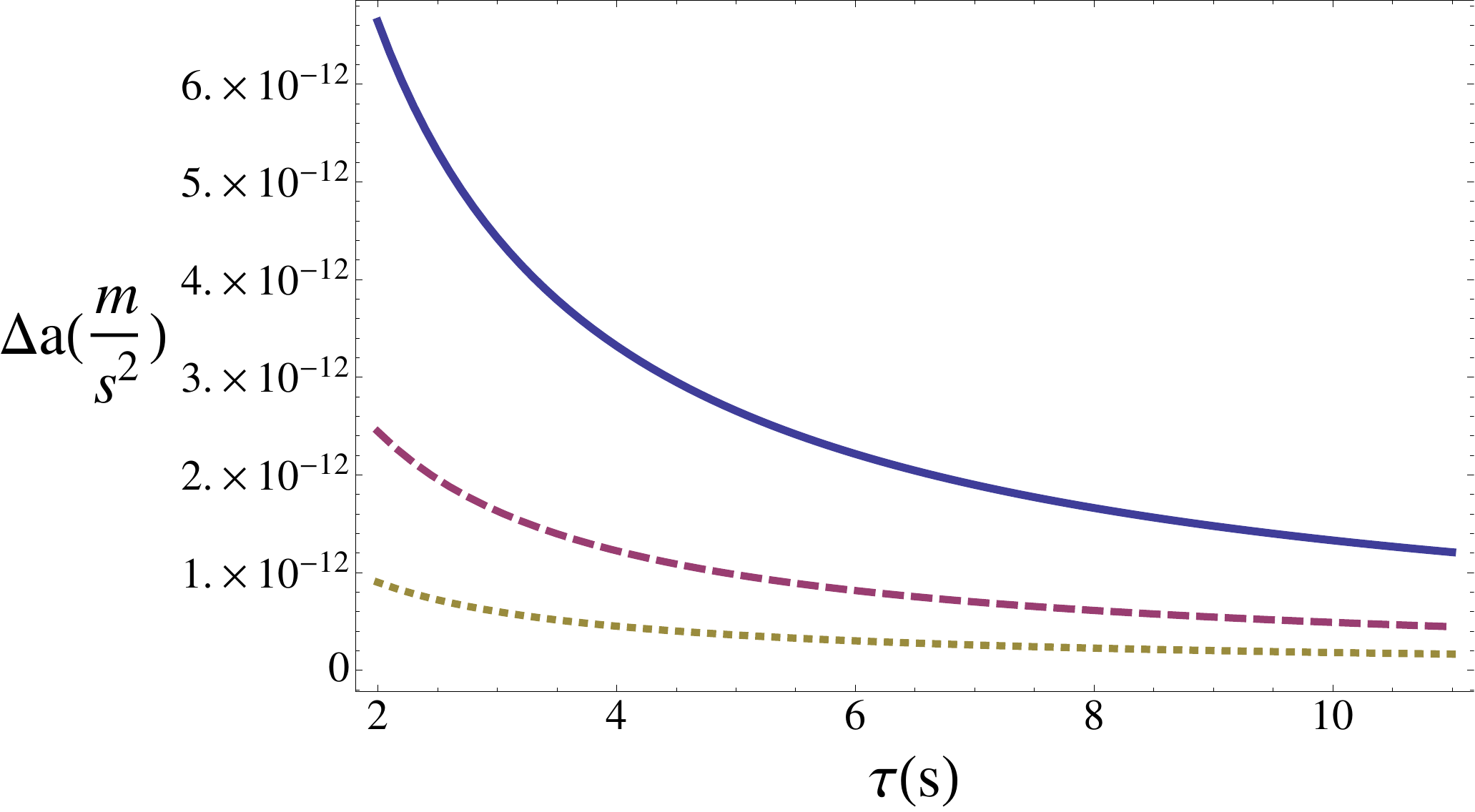}
\caption{Optimal bound $\Delta\,a (\operatorname{m/s^2})$ on the error in the estimation of the acceleration  vs proper time of acceleration $\tau (s)$ for squeezing parameter $r=8$ (blue, solid), 9 (red, dashed) and 10 (green, dotted). The frequencies of the modes are given by $\omega_n=2\pi\cdot 500\,n\,\operatorname{Hz}$, as corresponds to $L=1\operatorname{\mu\,m}$ and $c_s=10^{-3}\operatorname{m/s}$. The motion is assumed to be a sinusoidal oscillation of maximum acceleration $a$ and frequency $\omega=\omega_k+\omega_{k^{\prime}}$ and the number of measurements is $N=10^{11}$. The mode numbers are $k=1$ and $k^{\prime}=$ 2} \label{fig:results}
\end{figure}

We consider that the BEC undergoes sinusoidal acceleration given by $h(\tau)=h\,\operatorname{sin}(\omega\,\tau)$ and we use our scheme which exploits relativistic particle creation to estimate the amplitude $h$. For this continuous motion the Bogoliubov coefficients are given by  \cite{alphar},
\begin{eqnarray}
\alpha^{(1)}_{kk^{\prime}}(\tau)&=&i\,e^{-i\,\omega_{k^{\prime}}\tau} \alpha^{(1)}_{kk^{\prime}}(\omega_{k^{\prime}}-\omega_k)\int_0^{\tau} dt\operatorname{sin}(\omega\,\tau)e^{i\,(\omega_{k^{\prime}}-\omega_k)t}\nonumber\\
\beta^{(1)}_{kk^{\prime}}(\tau)&=&i\,e^{-i\,\omega_{k^{\prime}}\tau} \beta^{(1)}_{kk^{\prime}}(\omega_{k^{\prime}}+\omega_k)\int_0^{\tau} dt\operatorname{sin}(\omega\,\tau)e^{i\,(\omega_{k^{\prime}}+\omega_k)t}
\end{eqnarray}
where $ \alpha^{(1)}_{kk^{\prime}}=\frac{- 2 \sqrt{k\,k^{\prime}}}{\pi^2\, (k^{\prime}-k)^3}$ and $\beta^{(1)}_{kk^{\prime}}=\frac{2 \sqrt{k\,k^{\prime}}} {\pi^2\, (k+k^{\prime})^3}$ are the Bogoliubov coefficients that relate solutions to the Klein Gordon equation in the inertial and accelerated frames. It is easy to show that these coefficients reduce to the identity when $c^2\rightarrow\infty$ and in the Newtonian limit mentioned above. We assume that the frequency of the cavity oscillation is $\omega=\omega_k+\omega_{k^{\prime}}$ which generates a particle creation resonance. In Fig. (\ref{fig:results}) we plot the error in the estimation of the sinusoidal amplitude $h$ for typical experimental parameters. The QFI, given by Eq.~(\ref{fishresult}), for an initial squeezing of $r=10$, mode frequencies $\omega_k=2\pi\cdot 500\,\,\operatorname{Hz}$, $\omega_k^{\prime}=2\omega_k$, length $L=1\operatorname{\mu m}$ and effective velocity $c_s=10^{-3}\operatorname{m/s}$ is approximately $H(h)\simeq H(a)\simeq 10^{16}$. Note that our perturbative analysis is restricted to $a\ll 10^{-8}\operatorname{m/s^2}$.

It is interesting to compare our methods with techniques that have been previously developed to estimate accelerations using BECs within the framework of non-relativistic quantum mechanics. In particular we consider the QFI in state-of-the-art  accelerometers designed for the detection of gravity anomalies \cite{sagas,compact}, which are based on Mach-Zehnder atomic interferometry. In these schemes the wave function of the BEC is split and recombined using laser pulses, giving rise to a phase shift of $\phi=k\,a\,T^2$, where $k$ is the wave number of the atomic hyperfine transition, $a$ is the average acceleration and $T$ the interrogation time between pulses. The QFI in this case is given by \cite{nuestrotermo} $H=(\partial_a\,\phi)^2=(k\,T^2)^2$. Note that $k$ is fixed by the atomic species employed ($k=1.6\cdot 10^7 m^{-1}$ in Rb) and $T$ is limited by the dimensions of the experimental setup, being $1\, s$ in the best scenarios. Therefore, $H=2.6\cdot10^{14} \operatorname{m/s^2}$. The readout of the phase is then performed by fluorescence imaging of the atoms. The optimal sensitivity in the measurement of acceleration as provided by the QFI is given by \cite{sagas,compact}: $\delta a =1/(\sqrt{N} k T^2)$. After one measurement cycle, N is given by the number of detected atoms. After several cycles, the number of atoms is multiplied by the repetition rate and the integration time. Considering \cite{sagas,compact} a number of atoms of $10^6$---which already assumes a signal-to-noise ratio with respect to the total number of atoms in the BEC---a repetition rate of 5 Hz and integration time of a few hours, a number $N=10^{11}$ is obtained.  Therefore, those schemes require  a stable large-atom BEC machine \cite{becmachines}  in order to achieve large integration times. The absolute sensitivity in \cite{sagas,compact} is impressively small ($\Delta a\simeq 5\cdot 10^{-12}\,\operatorname{m/s^2}$).  In comparable conditions, our scheme which has been designed within relativistic quantum field theory is capable of improving the QFI by two orders of magnitude. Assuming that it is possible to realize the same number of measurements as done in the non-relativistic schemes, the optimal bound to the absolute sensitivity in our setup is around $1\cdot10^{-13} \operatorname{m/s^2}$, yielding a performance that goes several orders of magnitude beyond commercial accelerometers \cite{accelerometernatphot} ($\Delta a\simeq 10^{-4}\operatorname{m/s^2}$).

The performance of our scheme can be further improved by employing entangled two-mode initial states or by increasing the number of measurements. As can be seen in Fig.~(\ref{fig:results}), in our proposed implementation the error diminishes by increasing the degree of single-mode squeezing in the initial state. Notice that squeezing parameters of $r=10$ appear to be within reach for phonons in time-dependent potential traps \cite{squeezingserafini}. Furthermore, a large number of measurements can be in principle achieved by using atomic quantum dots or optical lattices coupled to the condensate in order to probe the state of the phononic field \cite{nuestrotermo}. Assuming as in \cite{nuestrotermo} that a few thousands of impurities can be coupled to the condensate and that each measurement can take a few milliseconds, the number of $10^{11}$ can be achieved after several hours of integration -that is, in conditions comparable to \cite{sagas,compact}. We note here that even with a much more modest level of squeezing such as $r=2$ and a  number of measurements of $10^4$, the predicted sensitivity of our device is $10^{-6}$, still two orders of magnitude beyond commercial accelerometers. Notice also that our setup is not restricted to a particular frequency of vibration, since there are resonances between any pair of oddly separated modes, both for particle creation and mode mixing. Therefore, a single BEC  is sensitive to several frequencies. Moreover, each BEC can be tuned at will by changing the length of the trap or the speed of the propagation of the phonons, using standard experimental techniques \cite{casimirwestbrook}. Taking all the above into account our setup can in principle exhibit a good broadband performance.

\section{Discussion}

The main aim of our research programme is the study of relativistic effects on quantum technologies. A comprehensive understanding of such phenomena will enable us not only to make the necessary corrections to technologies that are affected by them but also to use relativistic effects as resources. Indeed, we have shown that relativistic effects can be exploited to improve quantum precision measurements. In particular, we showed how particle creation within a moving cavity, a quantum field theoretical effect known as the dynamical Casimir effect, can be used to determine accelerations with a precision that, in principle, can improve state-of-the-art in accelerometers. As a particular experimental implementation, we have discussed a BEC setup. We showed that the QFI is several orders of magnitude larger than its counterpart in non-relativistic schemes. Therefore, the ultimate bound to the sensitivity of the accelerometer is several orders of magnitude smaller. This means that, by employing an optimal or close to optimal strategy of local phase estimation, our relativistic scheme can measure accelerations much smaller than the ones attained with optimal strategies in non-relativistic schemes. Moreover,  we have presented a general framework that can be used to measure parameters that appear in quantum field theory such as gravitational field strengths, proper times and accelerations.  Although Earth-based applications are also possible, the techniques are especially useful in space-based quantum technologies, where relativistic effects become relevant.

This paper establishes relativistic effects as resources in quantum technologies. Our work opens an avenue for the development of a new generation of relativistic quantum technologies.

\section*{Acknowledgements}
We warmly acknowledge Nicolai Friis for his contribution to early stages of this project. We thank Valentina Baccetti, Kai Bongs, Iacopo Carusotto, Jason Doukas, Marcus Huber, Antony Lee, Jorma Louko, Ralf Sch{\"u}tzhold, Augusto Smerzi and Angela White for useful discussions and comments.
M.~A., C.~S., and I.~F.  acknowledge support from the UK EPSRC [CAF Grant No.~EP/G00496X/2 to I.~F.]. D.~E.~B. acknowledges funding by the UK EPSRC  [Grant No.~EP/J005762/1 and hospitality from the University of Nottingham. G.~A. thanks the Brazilian funding agency CAPES [Pesquisador Visitante Especial-Grant No.~108/2012] and the Foundational Questions Institute [Grant No.~FQXi-RFP3-1317] for financial support.

\section*{Author contributions}
I. F. conceived the main idea and directed the project, M. A., D. E. B., and C. S. mainly executed the project, G. A. assisted in the methods development. All the authors contributed to the discussion and manuscript preparation.

\section*{Additional information}
The authors declare no competing financial interests. Correspondence and requests for materials should be addressed to I. F. (\verb"ivette.fuentes@nottingham.ac.uk").


\begin{thebibliography}{100}
\bibitem{footnote}
The analysis is also applicable to a continuous mode decomposition in the case that the Bogoliubov transformations mix a finite number of modes.
\end{thebibliography}

\end{document}